# Mechanical properties of hydrogen functionalized graphene allotropes

Yinfeng Li [1,a,*], Dibakar Datta [2,a], Zhonghua Li [1,*], Vivek B. Shenoy [3,4,*]

[a] *These authors made equal contribution*

[1] Department of Engineering Mechanics, Shanghai Jiao Tong University, Shanghai 200240, China
[2] School of Engineering, Brown University, Providence 02912, USA
[3] Department of Material Science and Engineering, University of Pennsylvania, Philadelphia PA 19104, USA
[4] Department of Mechanical Engineering and Applied Mechanics, University of Pennsylvania, Philadelphia PA 19104, USA

[*] Corresponding authors: Yinfeng Li (liyinfeng@sjtu.edu.cn); Zhonghua Li (zhli@sjtu.edu.cn), and Vivek B Shenoy (vshenoy@seas.upenn.edu)

**ABSTRACT**

Molecular dynamics (MD) simulations have been performed to investigate the mechanical properties of hydrogen functionalized graphene allotropes (GAs) for H-coverage spanning the entire range (0-100%). Four allotropes (graphyne, cyclic graphene, octagonal graphene, and biphenylene) with larger unit lattice size than graphene are considered. The effect of the degree of functionalization and molecular structure on the Young's modulus and strength are investigated, and the failure processes of some new GAs are reported for the first time. We show that the mechanical properties of the hydrogenated GAs deteriorate drastically with increasing H-coverage within the sensitive threshold, beyond which the mechanical properties remain insensitive to the increase in H-coverage. This drastic deterioration arises both from the conversion of $sp^2$ to $sp^3$ bonding and easy rotation of unsupported $sp^3$ bonds. Allotropes with different lattice structures correspond to different sensitive thresholds. The Young's moduli deterioration of fully hydrogenated allotropes can be up to 70% smaller than that of the corresponding pristine structure. Moreover the tensile strength shows an even larger drop of about 90% and higher sensitivity to H-coverage even if it is small. Our results suggest that the unique coverage-dependent deterioration of the mechanical properties must be taken into account when analyzing the performance characteristics of nanodevices fabricated from functionalized GAs.

1. **Introduction**

Over the last few decades, tremendous research attention has been devoted to the extraordinary electrical[1], mechanical[2] and thermal properties[3] of fullerene, nanotubes, and graphene. Recent advancement in the synthesis and assembly[4] process has led to the development of many new carbon materials. Particularly, the two-dimensional structures of carbon network with the same symmetry as graphene, such as carbine[5], graphane[6] and graphyne[7], have been extensively investigated experimentally[8, 9] and theoretically[10] due to their promising electrical[11] and optical mechanical properties[12, 13]. These two-dimensional graphene allotropes (GAs) can serve as precursors to build various nanotubes, fullerenes, nanoribbons, and other low-dimensional nanomaterials. Depending on their structural composition, the GAs can also be functionalized via chemical addition reaction in which the carbon atoms are converted from $sp^2$ to $sp^3$ hybrids to bond with the added chemical groups. Hydrogen adsorption on GAs has been acknowledged as an efficient way to modify their properties, such as tunable band gap[14], ferromagnetism[15] and thermal conductivity[16, 17]. The enhanced properties provided by hydrogenation are also tunable by changing the hydrogen coverage[18] [19-21].

To efficiently utilize such chemical functionalization with hydrogen atoms, it is necessary to understand the mechanical properties of the hydrogenated structures [22, 23]. Hydrogenation process causes membrane shrinkage and extensive membrane corrugation. This can lead to the deterioration of graphene mechanical properties[24, 25], such as Young's modulus, shear modulus and wrinkling properties. Many studies have been devoted to evaluate the mechanical properties of GAs using molecular dynamics (MD) simulations [26-28]. Pei *et al.* studied the effect of the degree of hydrogen coverage on mechanical properties of hydrogenated graphene and found that the Young's modulus, tensile strength, and fracture strain deteriorate drastically with increasing H-coverage. Their results suggest that the coverage-dependent deterioration of the mechanical properties must be taken into account when analyzing the performance characteristics of nanodevices fabricated from hydrogenated graphene allotrope sheets (GA sheets).[29] However, no investigation has been reported to date about the influence of hydrogen coverage on the mechanical properties of these GAs. Further research about the properties of their hydrogenated structures is in great need.

In this paper, we report the hydrogen coverage-dependent mechanical properties for graphyne and three new stable GAs.[30] Young's modulus and intrinsic strength of the chosen GAs are evaluated with varying H-coverage in the range of 0-100%. Moreover, the failure processes of some new GAs, such as Biphenylene, are reported for the first time. The mechanical properties of the investigated GAs deteriorate with increasing H coverage, and show different sensitivity to the functionalization. Our results suggest that novel failure mechanics is unique to a functionalized two-dimensional system.

1. **Model and method**

The atomic structures of the examined carbon networks are depicted in Fig. 1. The simulated allotropes have periodic boundary conditions with a lateral size of approximately 7.5nm. We arranged the carbon sheets by orienting armchair and zigzag edges along the X and Y axes, respectively. We used the LAMMPS [31] package for the MD simulations with an Adaptive Intermolecular Reactive Bond Order (AIREBO) potential [32] with an interaction cut-off parameter of 1.92 as used in the work of Pei *et al.* [29]. Prior to loading, the initial configurations were first relaxed to reach equilibrium. Tensile loading with a strain rate of 0.0005/ps was applied by displacing the simulation box followed by a relaxation for 10,000 MD steps. The time step of our simulations was 1fs. This procedure of relaxation and stretching was repeated for all the allotropes to evaluate their mechanical properties. All sets of the simulation were performed at room temperature under NVT ensemble. We first generated models of fully hydrogen functionalized GA sheets (H-100%) by bonding hydrogen atoms on one side of the carbon structure. Further hydrogenated GA sheets with certain H-coverage were achieved by randomly removing H atoms from fully hydrogenated hydrocarbon models.

The stress-strain curves during deformation can be obtained by following the study of Pei *et al.*[29] on the mechanical properties of hydrogen functionalized graphene. The atomic volume is taken from the initial (relaxed) sheet with the thickness of 3.4 Å. The stress of the carbon sheet is computed by averaging over all the carbon atoms in the sheet. From the simulated stress-strain curves, the Young's modulus $E$, ultimate strength $\sigma$ can be obtained. The Young's modulus is calculated as the initial slope of the stress strain curve and the strength is defined at the point where the peak stress is reached.

2. **Results and discussion**

The described MD approach is first verified by studying the mechanical properties of pristine graphene and hydrogen-functionalized graphene. The simulated Young's modulus and tensile strength of graphene are around ~0.86 TPa and 121GPa, which are in good agreement with experimental results of 1.0 TPa and 123.5 GPa, respectively. For the mechanical properties of graphene with different hydrogen coverage, our calculations show the same deterioration as reported by Pei *et al.*[29]. We also simulated the mechanical properties for GAs nanoribbons with a width of 7.5nm to test size effect on the simulated results. By comparing the results of nanoribbons with the case of the periodic boundary conditions, we find a very small change in the Young's modulus and tensile strength.

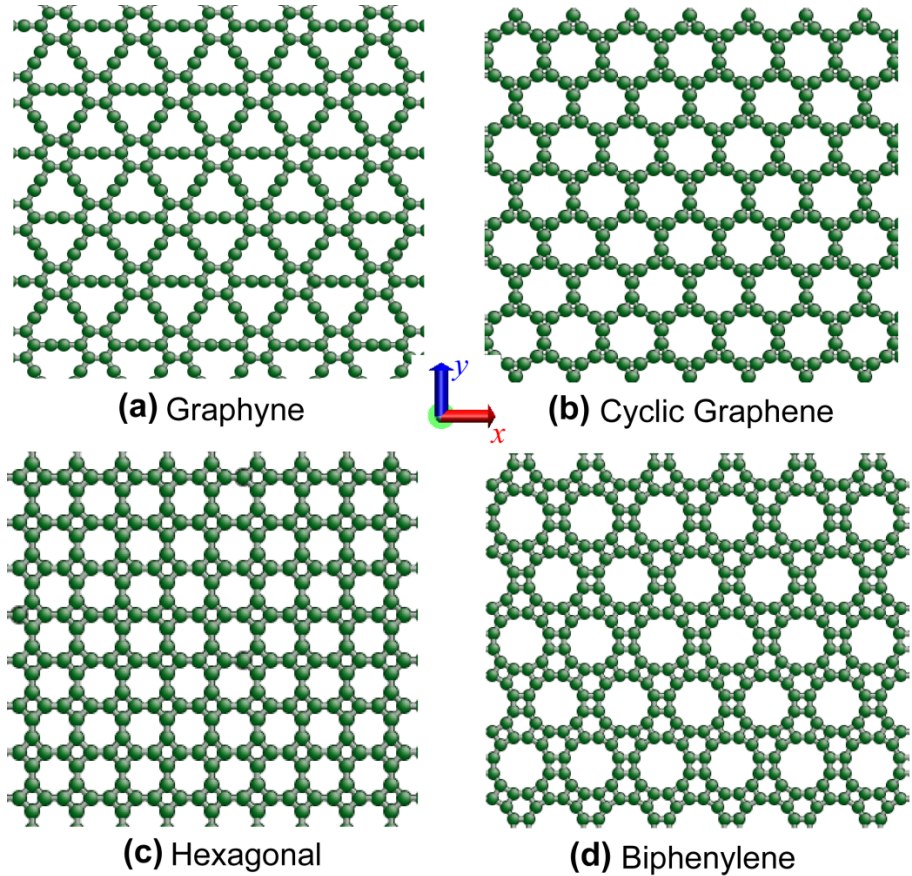

**Fig. 1.** Optimized pristine carbon atomic structures for the examined graphene allotropes.

We now proceed to study how mechanical properties are altered as GA sheets are functionalized with hydrogen. As shown in Fig. 1, we choose four different allotropes, which have been predicted by first-principles total energy calculations, including graphyne (benzene rings linked by diacetylene) as well as three stable 2D carbon supra crystals (biphenylene, cyclic and octagonal graphene). The chosen allotrope sheets are mixture of sp- and $sp^2$-hybridized carbon atoms network with the area of unit lattice being considerably larger than that of graphene. Compared to graphene, their larger surface areas allow a variety of potential applications for energy storage, such as hydrogen storage and lithium-ion batteries.[33, 34] Typical stress-strain curves of the functionalized allotropes with varying H-coverage are calculated. Fig. 2 shows the stress-strain curves of hydrogen functionalized Graphyne, Cyclic Graphene, Octagonal Graphene, and Biphenylene for H-coverage of 10%, 50% and 100%, together with that of the pristine allotrope sheet. The tensile strength of the investigated pristine GA sheets is much higher than pristine graphene implying their potential for wider applications. It can be seen that functionalized sheets fail at much lower stress and the corresponding fracture strain is also lower compared to that of the pristine sheet.

**Table 1** Simulated mechanical properties of pristine allotropes pulled in the X direction as in Fig. 1

| Allotropes | Young's modulus $E$ (TPa) | Strength $\sigma$ (GPa) | Fracture Strain $\varepsilon_F$ |
|---|---|---|---|
| Graphyne | 0.35 | 158.10 | 0.29 |
| Cyclic Graphene | 0.60 | 145.70 | 0.26 |
| Octagonal Graphene | 0.39 | 205.40 | 0.27 |
| Biphenylene | 0.43 | 162.10 | 0.28 |

The Young's modulus, tensile strength and fracture strain of the pristine GAs obtained from the stress–strain curves in Fig. 2 are given in Table 1.

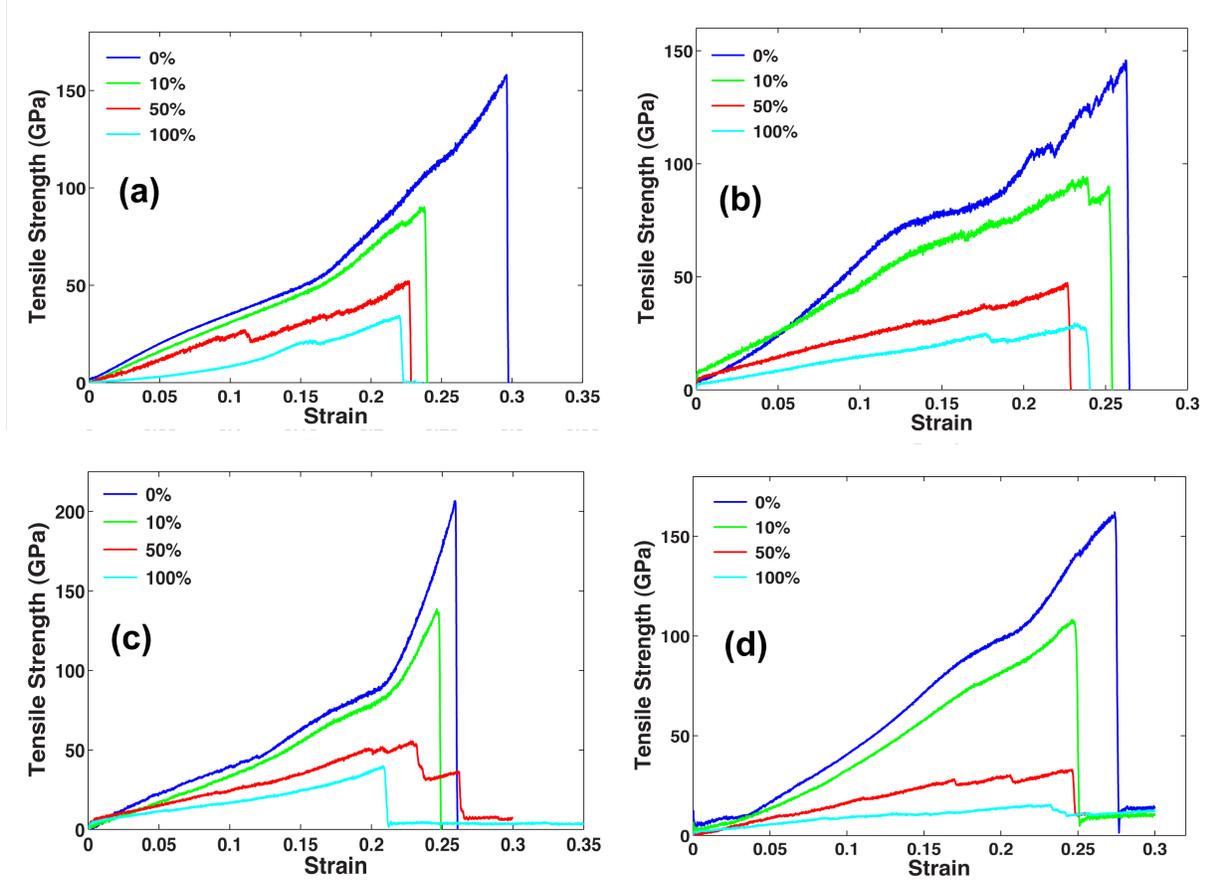

**Fig. 2.** Stress-strain curves of hydrogen functionalized (**a**) Graphyne (**b**) Cyclic Graphene (**c**) Octagonal Graphene and (**d**) Biphenylene for H-coverage of 0%, 10%, 50% and 100%.

The mechanical properties of the functionalized sheets obtained from extensive molecular dynamics simulations for different H-coverage are shown in Fig. 3. The strength and Young's modulus deteriorate because of the formation of weaker $sp^3$ bonds after H-functionalization. Local stress rearrangement induced by the conversion of local carbon bonding from $sp^2$ to $sp^3$ hybridization also contributes to the decay. The mechanical properties (Young's modulus $E$, tensile strength $\sigma$) for each hydrogenated allotrope are normalized by the properties of their corresponding pristine structure ($E_0$ and $\sigma_0$) as mentioned in Table 1. The error bars in the curves are obtained from simulations of statistically independent realizations of functionalized sheets for a given coverage.

As Fig. 3a shows, the Young's modulus of these four allotropes exhibits sharp decreases with the increase of hydrogen coverage from 0 to 50% with different sensitivities. The Young's modulus of cyclic graphene shows noticeable decay as H-coverage increases from 0 to 60%, beyond which the decay is not obvious. For graphyne, it reduces almost linearly to 30% of that of pristine as H-coverage increase from 0 to 100%. In case of cyclic graphene, there is sharp decay (almost 50%) until 30% H-coverage followed by about 15% more decay until 60% coverage. Beyond this percentage, however, there is no change in modulus. For octagonal graphene, two linear decay regimes can be identified in the range of 0-50% (50% decay) and 50-100% (10% more decay). The Young's moduli are less sensitive to the increase of hydrogen coverage after 90%. Fully hydrogenated biphenylene shows the largest decay (similar pattern like octagonal graphene) among the investigated allotropes.

The tensile strength $\sigma$ of these four allotropes shows coverage sensitive and insensitive regimes. It can be seen that the sensitive regime for cyclic graphene and graphyne is 0-40%. For octagonal graphene and biphenylene, it is 0-60%. In the sensitive regime, strength reduces by 65-70% for the first two allotropes while for the latter two cases, strength reduces by around 80%. In the coverage sensitive regime, the drop in strength is faster than the drop in Young's modulus. When H-coverage increases from 0 to 20%, the cyclic graphene shows lowest decay in tensile strength while graphyne shows fastest decay. It is interesting to notice that for the four allotropes considered, the order of decay speed in strength is opposite to that of the Young's modulus.

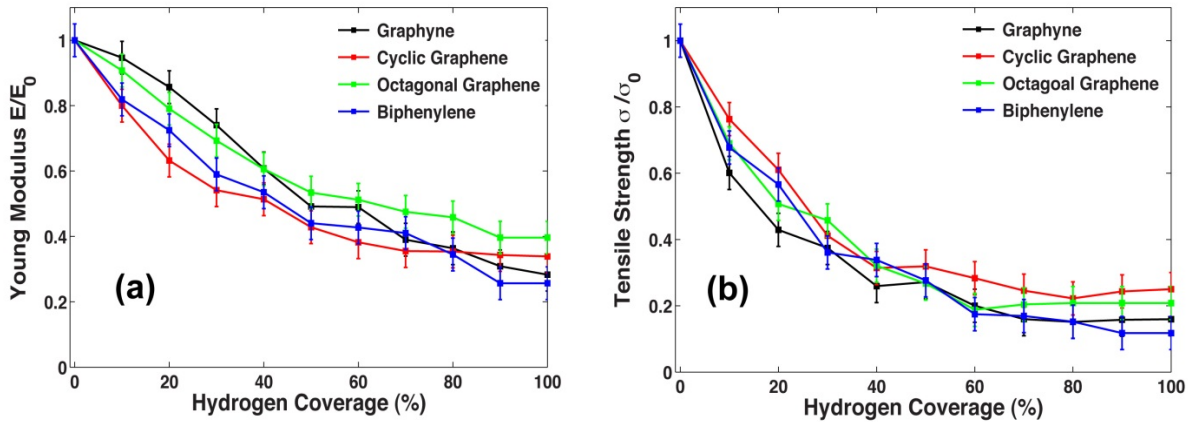

**Fig. 3**. Deterioration of (**a**) Young's modulus and (**b**) tensile strength for the investigated GAs for different H-coverage.

Fig. 4 shows a typical bond breaking, crack nucleation, and growth scenario in biphenylene functionalized with 40% H-coverage. The stress distribution on carbon atoms in the sheet before bond breaking is shown in Fig. 4b. It can be seen that the lattices deformed under external tension and the structure of benzene rings (hexagons $C_6$), which are interconnected by –C–C– linkages, sustain the largest stress. It is interesting to note that the bond breaking always initiates at the $sp^3$ bonds. Subsequently bonds outside the hydrogenated regions break leading to the crack formation (Fig. 4c), nucleation (Fig. 4d), propagation (Fig. 4e), and finally tearing of the sheet (Fig. 4f). We have also noted in all other allotropes that $sp^3$ bonds always break before the $sp^2$ bonds even when the latter bonds are subjected to larger stresses. This clearly shows that the $sp^2$ to $sp^3$ bonding transition results in the drop of mechanical strength of hydrogenated GAs. The failure occurs in the limit of large strain and is primarily controlled by breaking of weakest bonds, which can be triggered by even small H-coverage. This conclusion explains the noticeable drop of tensile strength when the hydrogen coverage increases from 0 to 40%. Since the Young's modulus measures an average deformation of the system in a small amplitude regime, its deterioration is less sensitive to hydrogenation than the decay of tensile strength, which has been demonstrated in Fig. 3. As the hydrogen coverage increases above the sensitive threshold, the stronger $sp^2$ bond network finally begins to disrupt. But weaker $sp^3$ bonds that make the mechanical properties insensitive to the H-coverage govern the failure behavior of the functionalized allotropes. Pei *et al.* reported the unconstrained rotation of $sp^3$ bonds caused by the stretching of the two dimensional graphene sheets. This unique phenomenon is also applicable for the investigated allotropes. The formation of unsupported $sp^3$ bonds leads to the reduction in the elastic modulus and the strength of graphene.

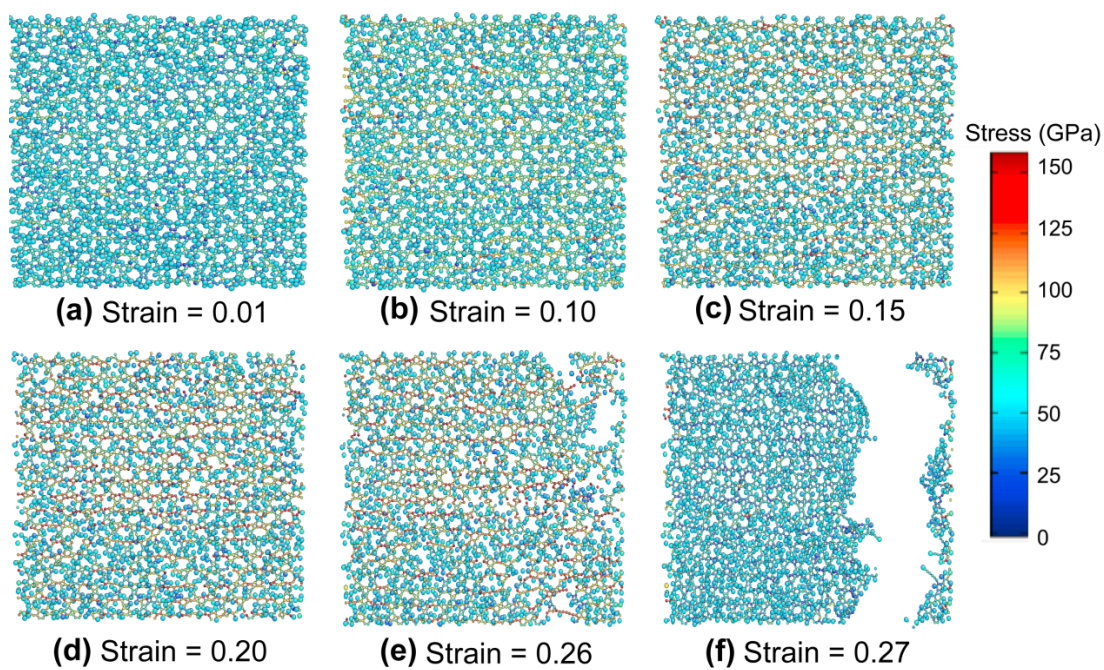

**Fig. 4.** Fracture in biphenylene sheet functionalized with 40% H-coverage: (**a**) configuration at initial stage (**b**) onset of bond breaking (**c**) crack formation, (**d**) nucleation, (**e**) propagation, and (**f**) tearing of the sheet.

## 3. Conclusion

In conclusion, we have carried out systematic MD simulations to study the mechanical properties of H-functionalized GAs with coverage spanning the entire range from 0-100%. Here we have found that the tensile strength is more sensitive to the increase of H-coverage than Young's modulus for all the allotropes, and it drops sharply even at small coverage. Among the chosen GAs, graphyne shows sharpest decay in tensile strength and lowest deterioration in Young's modulus as H-coverage increases from 0 to 20% because of its lowest surface density. Different allotropes exhibit different deterioration pattern and sensitive regimes, which can provide guidance for the potential application for hydrogen storage. Our simulations also show that bond breaking always initiates at $sp^3$ bonds even with the presence of hybrid $sp^2$ bonds, and therefore the $sp^3$ bonding transition contributes to the loss of strength of functionalized GAs. Our results suggest that the coverage-dependence of the mechanical properties should be taken into account in analyzing the performance characteristics of mass sensors, nanoresonators, and impermeable membrane structures fabricated from functionalized GA sheets. In this paper, we only focused on the decay trend and mechanism induced by hydrogen functionalization. Further analysis is expected into the dependence of mechanical and other properties on stretch directions and hydrogen arrangement pattern on the surface of carbon sheets. The experimental verifications of our results are also expected.

## Acknowledgements

We gratefully acknowledge the support of National Science of Foundation (USA), Department of Energy (USA), and the National Basic Research Program of China (No. 10932007). The computational support for this work was provided by grant TG-DMR090098 from the TeraGrid Advanced Support Program.

**Figure Legends**

**Fig. 1**. Optimized pristine carbon atomic structures for the examined graphene allotropes.

**Fig. 2**. Stress-strain curves of hydrogen functionalized (**a**) Graphyne (**b**) Cyclic Graphene (**c**) Octagonal Graphene and (**d**) Biphenylene for H-coverage of 0%, 10%, 50% and 100%.

**Fig. 3**. Deterioration of (**a**) Young's modulus and (**b**) tensile strength for the investigated GAs for different H-coverage.

**Fig. 4**. Fracture in biphenylene sheet functionalized with 40% H-coverage: (**a**) configuration at initial stage (**b**) onset of bond breaking (**c**) crack formation, (**d**) nucleation, (**e**) propagation, and (**f**) tearing of the sheet.

13